\documentclass[11pt,a4paper,fleqn]{article}
\usepackage[english]{babel}
\usepackage{epsfig}
\usepackage[utf8]{inputenc}
\usepackage{floatrow}
\usepackage{amssymb,amsmath,amsfonts,amsthm,graphicx,psfrag}
\usepackage{indentfirst}
\usepackage{hyperref}
\usepackage[title,titletoc]{appendix}
\usepackage{bm}
\usepackage{verbatim}
\usepackage{epsfig}
\graphicspath{{./pictures/}}
\usepackage{authblk}
\usepackage{cite}
\usepackage[title,titletoc]{appendix}
\usepackage{slashed, braket}
\usepackage[left=2cm,right=2cm,
    top=2cm,bottom=2cm,bindingoffset=0cm]{geometry}
		\newcommand{\be}{\begin{eqnarray}}
\newcommand{\ee}{\end{eqnarray}}
\newcommand{\ba}{\begin{aligned}}
\newcommand{\ea}{\end{aligned}}
\usepackage{authblk}

\newcommand{\tr}{\text{tr }}
\newcommand{\Tr}{\text{Tr }}
\newcommand{\Det}{\text{Det }}
\newcommand{\al}{\alpha}

\begin{document}

\title{Chiral separation effect for spin 3/2 fermions.}

\author{Z.V.Khaidukov\\ e-mail: \href{khaidukov.zv@phystech.edu}{khaidukov.zv@phystech.edu}
    \and R.A.Abramchuk\\ e-mail: \href{abramchuk@phystech.edu}{abramchuk@phystech.edu}}
\affil{Moscow Institute of Physics and Technology, 9, Institutskii per., Dolgoprudny, Moscow Region, 141700, Russia}
\affil{Institute for Theoretical and Experimental Physics of NRC ``Kurchatov Institute'', B. Cheremushkinskaya 25, Moscow, 117259, Russia}

\maketitle

\begin{abstract}
    Chiral Separation Effect (CSE) for systems that feature spin 3/2 fermions was considered.
    For the self-consistent Adler's model with relativistic massless Rarita-Schwinger fermions (RSA model), we found that the CSE conductivity is five times larger than for massless Dirac fermions.
    For a model of four-fold band crossing in Rarita-Schwinger-Weyl semimetals, in which massless fermions with quasispin 3/2 exist, we calculated that the CSE conductivity is four times larger than for Weyl fermions.
    We show that CSE conductivity for any multi-degenerate Fermi point in topological semimetals is proportional to its Chern number and is topologically protected.
    Along the calculations, we proved an index theorem that relates Chern number of a Fermi-point and spectral asymmetry of the corresponding Landau band structure.
    The assumption that CSE for any system of chiral fermions is dictated by the corresponding Chern number is found to be correct for RSA model (and for the Dirac fermions).
\end{abstract}

\begin{section}{Introduction}
The chiral separation effect (CSE) emerges in a dense medium of charged fermions in external magnetic field.
Resulting axial current density \(\mathbf{j}^{5}\) 
\footnote{Vector and axial currents are introduced as $j^\nu=\bar\psi\gamma^\nu\psi=j^\nu_{l}+j^\nu_{r}$, $j^{5\nu}=\bar\psi\gamma_5\gamma^\nu\psi=j^\nu_{l}-j^\nu_{r}$. } 
is  directed along the magnetic field \(\mathbf{B}\).
For massless Dirac fermions with electric charge \(e\), the well-known result is \cite{Metl} (in this paper, we denote CSE conductivity with \(\sigma\))
\be
    &&\mathbf{j}^{5}=\sigma_D\mathbf{B},\quad \sigma_D = \frac{e}{2\pi^{2}}\mu, \label{stdCSE}
\ee 
where \(\mu\) is the chemical potential.  
   
From the theoretical point of view, CSE is distinctive due to the following features.
Firstly, the effect exists in thermodynamic equilibrium --- the resulting current is non-dissipative. 
Though the original derivation \cite{Metl} employed subtractions of non-renormalized quantities
\footnote{For the Chiral Magnetic Effect (CME) \cite{Kharzeev1,Kharzeev2} such a subtraction leads to an incorrect result. 
    The problem of regularization in fundamental theory is associated with the problem of vector/axial currents conservation. }
, analytical studies \cite{KZ2017} in a well-regularized theory and numerical studies \cite{PBCSE} confirmed the result.
    
Secondly, CSE conductivity is insensitive to a wide class of interactions. 
In lattice QCD simulations immutability of the coefficient was shown \cite{PBCSE}, 
while in analytical studies of theories with four-fermion interaction corrections were found \cite{NJL1, NJL2, NJL3, Holo1, Holo2}. 

A common theoretical argument for CSE persistence to interactions is its relation to the chiral anomaly, which is protected from perturbative corrections. 
The relation may be demonstrated with a mnemonic derivation: the substitution
$ \mu \to \mu + eA_0 $ in \eqref{stdCSE}, and divergence applied to the both sides yields (let the fields be directed along z-axis)
\be 
    &&\mathbf{E}=-\nabla A_0,~\partial_{z}\mu = 0\to  
    \quad\partial_{z} j^{5z} =  \frac{e^{2}}{2 \pi^{2}}{EB}
\ee

    Another theoretical argument, which was proposed in the original paper \cite{Metl}, is the topological nature of the effect for massless fermions.
CSE conductivity is immutable in any interacting theory of massless fermions that is topologically equivalent to the free theory.
For instance, in QCD with massless quarks.

From the experimental perspective, the effect may be independently observed in high energy and in solid state physics. 
Dense medium of quarks emerges in high energy Heavy Ion Collisions, where the light quarks masses are negligible.
The massless quasiparticles in crystals emerge due to spectra linearity near the conductivity bands crossing points (Fermi points, or nodes), which stability is of topological nature.
    
Chiral separation and magnetic effects can be spawned by effective gauge fields, including emergent gravitational fields, which is especially natural for condensed matter physics.
In general, the chiral effects are argued to be related to axial-gravitational anomalies \cite{semimetal_effects7}. 
Effective gauge fields are spawned by deformations and dislocations in crystals, and by overall motion, e.g.~rotation, in any medium.
A dislocation spawns effective electric field \cite{eqBerry}.
Twisting spawns the skew-symmetric part of the metric tensor, which yields the Chiral Torsional Effect \cite{CTE,CTE2,CTE3,CTE4,CTE5,CTE6}.
A rotation spawns an effective uniform (near the rotation axis) magnetic field, which yields the Chiral Vortical Effect \cite{Chern,Vilenkin,AKZ}.

In the present paper, we calculate the Chiral Separation Effect for systems of massless fermions with various momentum space topological charges, or Chern numbers.
As a relativistic model, we study Adler's extension \cite{Adler1,Adler2,Adler3} of the Rarita-Schwinger equation \cite{RS}, and find that \(\sigma_{RSA}=5\sigma_D\).
Then we study the effect in an effective theory for so-called  Rarita-Schwinger-Weyl semimetals (RSW) \cite{RSW}, and find that \(\sigma_{RSW}=4\sigma_W\)
    (where \(\sigma_W=\sigma_D/2\) is the conductivity associated with an isolated Weyl point).
Massless charged (quasi)spin 3/2 quasiparticles emerge due to fourfold band crossings
    (while a Weyl point emerges due to a two-fold band crossing),
which were experimentally discovered in a number of crystals, e.g. CoSi, RhSi \cite{RSexp1,RSexp2,RSexp3}, AlPt \cite{RSexp4}, PdGa \cite{Schroter:2020}, PdBiSe \cite{RSexp5}.
    For a multi-fold band crossing node, we find that the CSE conductivity \(\sigma_s\) is dictated by the node's Chern number \(\mathcal C_s\) 
    (the subscript \(s\) labels a type of the node, as explained in section \ref{SSCSE};
    in the context of topological semimetals, we assume that a quasiparticle is of unit charge \(e=1\))
\be
\sigma_s = \frac{\mu}{4\pi^2}\mathcal C_s.
\ee
Application of this result to RSA model explains the difference between \(\sigma_{RSA}\) and \(\sigma_{RSW}\).
Along the calculations, we (re)discover an index theorem \cite{LandauIndex}, which relates the node's Chern number and spectral asymmetry of the corresponding Landau band structure.

The paper is organized as follows.
In section \ref{RSA}, RSA model is reviewed. 
In section \ref{CSE32}, CSE for RSA model is calculated, relation to the anomaly is discussed, and, in subsection \ref{CSEstructure}, structure of the result is discussed. 
In section \ref{SSCSE}, the effect in RSW semimetals is studied.
In subsection \ref{CSEexact}, the result is generalized to other types of Fermi-points, and the index theorem is derived. 
Section \ref{SectConc} contains conclusions and discussions. 
  
\end{section}

\begin{section}{Rarita-Schwinger equation  and the Adler's extension}
\label{RSA}

    The original equation was formulated in 1941 by Rarita and Schwinger to describe hypothetical charged relativistic fermions with spin 3/2. 
    The corresponding action
    \footnote{Mind that we use the metric convention \(\eta_{\mu\nu} = (+,-,-,-)\), and the corresponding gamma matrices.}
    involves a four-vector of four-spinors \(\psi_\mu\) -- the Rarita-Schwinger (RS) field, and the electromagnetic field \(A_\mu\)
\be
    &&S_{RS}[\psi_\mu, A_\mu]=\int d^4x~\bar{\psi}_{\mu}\epsilon^{\mu\nu\lambda\rho}\gamma_{5}\gamma_{\nu}iD_{\lambda}\psi_{\rho},
    \quad 
    D_{\nu}\psi_{\rho}=(\partial_{\nu}-ieA_{\nu})\psi_{\rho},\\
    &&\bar\psi_\mu = \psi_\mu^\dag\gamma_0,~
    \gamma_0=\gamma^0,~
    \gamma_5=i\gamma_0\gamma_1\gamma_2\gamma_3 = \text{diag}(-1,-1,1,1).
\ee

    In the absence of the gauge field, the action possesses an additional symmetry
\be
    &&\psi_{\rho} \to \psi_{\rho}+\partial_{\rho } \epsilon,  \label{SpinGaugeSym}
\ee 
    where $\epsilon$ is an arbitrary four-spinor. 
    The symmetry transformation is remarkably similar to the standard Abelian vector field gauge transformation.
    The symmetry implicates, formally, that the Green function for the free massless RS equation, or the propagator of \(\psi_\mu\), doesn't exist.
    The standard in perturbative QFT resolution would be to introduce the gauge-fixing term to the Lagrangian 
    \(L_\zeta = \zeta \bar{\psi}_{\mu}\gamma^{\mu}\gamma^{a}\partial_{a} \gamma^{\rho}\psi_{\rho}\).
    
    Since the formulation of the RS equation until very recently, all the attempts to introduce a gauge interaction failed.
    The interacting theories, e.g.~minimally coupled Abelian vector gauge field, suffer from inconsistencies.
    At the classical level, the theories feature solutions that propagate with superluminal speed \cite {6}.
    At the quantum level, the equal time commutation relations and relativistic covariance are not compatible \cite{5}.
    And classical, and quantum considerations are beset by the first-class constraint \(\gamma_\mu\psi^\mu=0\) \cite{7,8}.
    
    A way to consistently introduce the gauge interaction was recently proposed by Adler \cite{Adler1,Adler2,Adler3}. 
    The theory contains an additional Dirac spinor field \(\lambda\) directly coupled to the RS field with dimensionfull constant \(m\) 
\be
    &&S = S_{RS} + S_{D} + S_{A},\\
    &&S_D[\lambda, A_\mu] = \int{d^4x~\bar{\lambda}\gamma_{\nu}iD^{\nu}\lambda},
    \quad D_{\nu}\lambda=(\partial_{\nu}-ieA_{\nu})\lambda,~
    \bar{\lambda}=\lambda^{\dagger}\gamma^{0},\\
    && S_A[\psi_\mu, \lambda] = -m\int d^{4}x~(\bar{\lambda}\gamma^{\nu}\psi_{\nu}-\bar{\psi}_{\nu}\gamma^{\nu}\lambda).
\ee
    In the extended theory, the malignant constraint, which is of second class, has been resolved by Adler. 
    Moreover, the extension breaks symmetry \eqref{SpinGaugeSym}.

    The constraint introduces an essentially new complication to calculations --- ghost degrees of freedom.
    In the anomaly computation \cite{Adler1} ghosts were considered in several ways that lead to 
    \begin{enumerate}
        \item non-propagating ghosts, which does not contribute  to the anomaly;   
        \item ghosts of infinitely large mass $\sim \lim_{\delta \to 0} \frac{m}{\delta}$, with -1 contribution; 
        \item exclusion of contribution 1 in Alvares-Gaume-Witten's manner \cite{GaumeWitten}.
    \end{enumerate}
    Besides, In \cite{Adler3} a model with additional auxiliary fields was investigated, and the same answers as for non-propagating ghosts were found.

    To be able to perform analytical calculations, we set $m\to \infty$. 
    Then the  Feynman rules read (for derivation see \cite{Adler1}) in the basis of Fourier-transformed fields \((\psi_\mu(k), \lambda(k))\)
\be
    &&N=\begin{pmatrix}N^{\frac{3}{2}}_{\rho\sigma}&0\\ 0 & 0 \end{pmatrix},\quad
    N^{\frac{3}{2}}_{\rho \sigma}=\frac{-i}{2k^{2}}(\gamma_{\sigma}\not{k}\gamma_{\rho}-\frac{4}{k^{2}}k_{\rho}k_{\sigma}\not{k}),\\
    &&V^{\nu}=\begin{pmatrix}-ie\gamma^{\mu\nu\rho}&0\\ 0 & -ie\gamma^{\nu} \end{pmatrix},\quad
    A^{\nu}=\begin{pmatrix}-i\gamma^{\mu\nu\rho}\gamma_{5} & 0 \\ 0 & -i\gamma^{\nu}\gamma^{5}  \end{pmatrix}, \\
    &&\gamma^{\mu\nu\rho}=\frac{1}{2}(\gamma^{\mu}\gamma^{\nu}\gamma^{\rho}-\gamma^{\rho}\gamma^{\nu}\gamma^{\mu}).
\ee   
    \(N\) is the propagator,  $V^{\nu} $ and $A^{\nu}$ --- the vector and axial vertices.
    The auxiliary field \(\lambda\) is non-propagating (\((N)_{22} \equiv \braket{\lambda\bar\lambda} = 0\)), yet interacting, as well as the ghosts.

    At this point, we have all the elements to derive the chiral conductivity.
\end{section}

\begin{section}{Chiral separation effect for Rarita-Schwinger-Adler model}
\label{CSE32}

    We calculate the conductivity within the linear response theory \cite{Landsteiner:2012kd,Kharzeev:2009pj}
\be
    &&\sigma=\lim_{p_{i} \to 0} \lim_{\omega \to 0}\frac{i}{2 p_{i}}\epsilon^{ijk}\Pi^{AV}_{jk}(p),
    \quad
    \Pi^{AV}_{jk}=\int d^{4}x~ e^{ip^{\mu}x_{\mu}}\braket{j^{5}_{j}j_{k}},
\ee
    where $j_{j}(x)$ and $j^{5}_{k}(x)$ are spatial components of the vector and axial currents, correspondingly, and \(\omega\equiv p_0\).

    With the Feynman rules from the previous section 
\be
&&\sigma_{RSA}=\lim_{p_{i} \to 0} \lim_{\omega \to 0}\frac{i\epsilon^{ijk}}{2 p_{i}}T\sum_n e^{2}\tr\left.\int{\frac{d^{3}r}{(2\pi)^{3}}N_{\rho \eta}(r)A_{j}^{\rho \eta }N_{\eta \mu }(r+p)V_{k}^{\mu\rho}}\right|_{r_0 \to i\omega_{n}+\mu}. \label{Kubo}
\ee
    Chemical potential and temperature are introduced with the Matsubara's prescription $k_{0}+i0\to i\omega_{n} + \mu$, 
    where \(\omega_n = 2\pi T(n+\frac12)\) are the fermionic frequencies.

    With the following algebraic relations
\be
    &&\gamma^{\rho}N_{\rho\sigma}=0,~
    N_{\rho\sigma}\gamma^{\sigma}=0,\quad
    \gamma^{\mu\lambda\rho}=\frac{1}{2}[\gamma^{\mu},\gamma^{\lambda} ]\gamma^{\rho}-\gamma^{\mu}g^{\lambda \rho}+g^{\mu \rho} \gamma^{\lambda}
,\\
    &&\gamma_{\mu}\gamma^{\tau}\gamma^{\rho}\gamma^{\sigma}\gamma^{\nu}\gamma^{\mu}=2\gamma^{\sigma}\gamma^{\rho}\gamma^{\tau}\gamma^{\nu}+2\gamma^{\nu}\gamma^{\tau}\gamma^{\rho}\gamma^{\sigma},\quad
    \gamma_{\mu}\gamma^{\tau}\gamma^{\rho}\gamma^{\mu}=4g^{\tau\rho},\quad\gamma^{\mu}\gamma_{\nu}\gamma_{\mu}=-2\gamma_{\nu},
\ee 
    we obtain
\be
    && N_{\alpha \beta}(r)\gamma^{j}\gamma^{5}N_{\beta \alpha}(p+r)\gamma^{k}
    = \left(1+4\frac{(r\cdot(r+p))^{2}}{r^{2}(r+p)^{2}}\right)\frac{\tr(\not{r}+\not{p})\gamma^{k}(\not{r})\gamma^{j}\gamma^{5}}{r^{2}(r+p)^{2}}.\label{eqA}
\ee

The first term of \eqref{eqA} inserted in \eqref{Kubo} yields \cite{KZ2017} 
\be
    \sigma^{(1)}_{RSA}&=&\lim_{p_{i} \to 0} \lim_{\omega \to 0}\frac{i\epsilon^{ijk}}{2 p_{i}}T\sum_{n} \left.\int{\frac{d^{3}r}{(2\pi)^{3}}\frac{\tr(\not{r}+\not{p})\gamma^{k}(\not{r})\gamma^{j}\gamma^{5}}{r^{2}(r+p)^{2}}}\right|_{r_0 \to i\omega_{n}+\mu} \\
    &=&\frac{e\mu}{2\pi^{2}} \label{cse1},
\ee
which is equal to the ordinary CSE conductivity.

The second term yields
\be
    \sigma^{(2)}_{RSA}=\lim_{p_{i} \to 0} \lim_{\omega \to 0}\frac{i\epsilon^{ijk}}{2 p_{i}}T\sum_{n} \left.\int{\frac{d^{3}r}{(2\pi)^{3}}~4\frac{(r\cdot(r+p))^{2}}{r^{2}(r+p)^{2}}\frac{\tr(\not{r}+\not{p})\gamma^{k}(\not{r})\gamma^{j}\gamma^{5}}{r^{2}(r+p)^{2}}}\right|_{r_0 \to i\omega_{n}+\mu}. 
\ee
Since convergence of the integral is questionable, we reduce the problem to the previous one \eqref{cse1} 
by means of a simple subtraction 
\be
    \sigma^{(2)}_{RSA} = (\sigma^{(2)}_{RSA}-4\sigma^{(1)}_{RSA})+4\sigma^{(1)}_{RSA}.
\ee
The integrand in the parenthesis
\be
    4\left(\frac{(r\cdot(r+p))^{2}}{r^{2}(r+p)^{2}}-1\right)
    \frac{\tr(\not{r}+\not{p})\gamma^{k}(\not{r})\gamma^{j}\gamma^{5}}{r^{2}(r+p)^{2}} = 
    4\frac{(r \cdot p)^{2}-r^{2}p^{2}}{r^{2}(r+p)^{2}}~
    \frac{\tr(\not{r}+\not{p})\gamma^{k}(\not{r})\gamma^{j}\gamma^{5}}{r^{2}(r+p)^{2}},
\ee
yields an obviously convergent integral, which is zero at \(p_i\to 0\).

The divergence of the sum over Matsubara frequencies rises from the `usual' divergent constant in the free energy, which is to be consistently dropped since it is independent of temperature or momentum.
For simplicity, let us remind how the difficulty is circumvented \cite{Le-Bellac} for Dirac fermions
\be
&&Z=\int D\bar{\psi}D\psi \exp\left(-\int d\tau d^{3}x ~\bar\psi(\tau,x)\slashed{D}\psi(\tau,x)\right),
\quad\slashed{D}= \gamma_{\nu}(i\partial_{\nu}-\mu\delta_{\nu 0}),\\
&&\Omega=-T\log Z= -T \log\Det{\slashed{D}}=-\frac{T}{2} \Tr\log(D^{2})=\sum_{k}\Omega_{k},
\quad   \Omega_{k}=\frac{T}{2}\sum_{n}\log(\omega_{n}^{2}+\omega_{k}^{2}).
\ee 
Though the free energy \(\Omega_k\) is divergent, its derivative with respect to \(\omega_k\) is well-defined and may be integrated back to a finite result (plus the constant)
\be
\frac{1}{2\omega_{k}}\frac{d\Omega_{k}}{d\omega_{k}}=\frac{T}{2}\sum_n\frac{1}{(\omega_{n}-i\mu)^{2}+\omega^{2}_{k}},
\ee
while we encountered its derivative with respect to \(\mu\)
\be
\frac{1}{2\omega_{k}}\frac{d^2\Omega_k}{d\mu d\omega_{k}}=-iT\sum_{n}\frac{\omega_{n}-i\mu}{((\omega_{n}-i\mu)^{2}+\omega_{k}^{2})^{2}} \label{equ12}.
\ee
In this way, CSE conductivity is related to the derivative of the free energy, which is well-defined as long as thermodynamics for the system exist.

Finally, the CSE conductivity in constant magnetic field for RSA model to the leading order in momentum reads
\be
    \sigma_{RSA} &=& \sigma^{(1)}_{RSA}+\sigma^{(2)}_{RSA}=5~\frac{e\mu}{2\pi^{2}} = 5\sigma_D.
\ee

The perturbative expressions at finite \(m\) are quite complicated.  
Per se, additive corrections \(\sim m^{-2},\,m^{-4}\) to the result are possible.
However, topological properties of the theory in the limit \(m\to\infty\) seem to preserve, 
which is an argument for the result to hold for any \(m>0\), as discussed in the next subsection. 

Besides, the calculation may be incomplete, unless the theory constraints are considered.
We argue that the ghosts does not contribute to CSE.
Firstly, the ghosts mass is proportional to \(\sim \lim_{\delta \to 0} \frac{m}{\delta}\to\infty\).
Heavy particles are suppressed by the thermodynamic distribution, hence their contribution to any observable is vanishing.
Secondly, following Zakharov's arguments \cite{SZ}, in the absence of interaction between particles, implicate formal equivalence of chemical and electrostatic potentials. 
Thus, the substitution $\mu \to \mu+eA_0(z)$ reproduces the anomaly, and demonstrates equivalence of the coefficient in anomaly and CSE conductivity,
though, an interaction might yield chemical potential renormalization and Fermi surface deformation. 

\subsection{On CSE attribution to spins in RSA model}
\label{CSEstructure}

RSA model is a theory of {\it combined} spin 3/2 Rarita-Scwinger and spin 1/2 Dirac fields \cite{Adler1}.
Thus, attribution of \(\sigma_{RSA}\) to spin 3/2 solely may be incorrect.

Let us approach the problem with the result of section \ref{SSCSE}.
According to that, CSE conductivity for a chiral system is dictated by its topological properties --- by Chern number \(\mathcal C\) for zero-energy node of the theory
\be
&&\sigma=\mathcal C\sigma_D, 
\quad\mathcal C \equiv \mathcal N_3 = \frac{1}{48\pi^2}\tr\gamma_5\int GdG^{-1}\wedge GdG^{-1}\wedge GdG^{-1},
\ee
where the integration is over a 3d-surface (in 4d Euclidean space) that embraces poles (and zeros) of the Euclidean propagator \(G\).
We suggest that \(\mathcal C\) may be expressed via helicities \(\lambda_\al\) of the left-chiral plane wave positive-energy solutions as \(\mathcal C=\sum_\al2\lambda_\al\).
The RSA model for \(m>0\) (including \(m\to\infty\)) has three such modes \cite{Adler1}: 
\(v_1\) and \(v_2\) are pure RS field modes with helicities \(\frac12\) and \(1+\frac12\), respectively,
and a `combined' mode \(v_3\) with helicity \(1-\frac12\).
Thus, \(\mathcal C_{RSA}=1+3+1\), and \(\sigma_{RSA}=5\sigma_D\).
Free massless RS theory features only maximum helicity modes, so \(\mathcal C_{RS}=3\), while the Adler's extension `defrosts' the \(\lambda=\frac12\) mode and adds the combined mode.

\end{section}

\begin{section}{CSE in Rarita-Schwinger-Weyl semimetals}
\label{SSCSE} 

    The content of the previous section may be of fundamental interest, yet is purely abstract.
    Meanwhile, condensed matter systems with dynamically induced relativistic invariance provide a playground for high energy physics.
    In this section, we calculate CSE in an effective theory for RSW semimetals \cite{RSW}, and generalize the result to other types of Fermi points, 
    in which charged quasiparticles with quasispin 3/2 exist.

    The effective theory is defined by the Hamiltonian in the standard basis \(\ket{s, s_z} \), \(s=\frac32\)
    \begin{gather}
H=v_f\sum_{i=1}^3\hat p_{i}S_{i} \label{eq1},\\
S_{1}=\begin{pmatrix} 0& \frac{\sqrt{3}}{2}&0&0\\
\frac{\sqrt{3}}{2}&0&1&0\\ 
0&1&0&\frac{\sqrt{3}}{2}\\
0&0&\frac{\sqrt{3}}{2}&0\\
\end{pmatrix},~ 
S_{2}=\begin{pmatrix} 0& \frac{-i\sqrt{3}}{2}&0&0\\
i\frac{\sqrt{3}}{2}&0&-i&0\\
0&i&0&-i\frac{\sqrt{3}}{2}\\
0&0&i\frac{\sqrt{3}}{2}&0\\
\end{pmatrix}, ~
S_{3}=\begin{pmatrix} \frac{3}{2}& 0&0&0\\
0&\frac{1}{2}&0&0\\
0&0&-\frac{1}{2}&0\\
0&0&0&-\frac{3}{2}\\
\end{pmatrix}
    \end{gather}

    Plane wave solutions of the corresponding Schrodinger equation read 
    \begin{gather}
        \psi_{\mathbf p\lambda}(\mathbf x, t) = u_{\mathbf p\lambda}\exp(-iE_{\mathbf p\lambda}t + i\mathbf p\mathbf x), 
        ~E_{\mathbf p\lambda} = v_f\lambda |\mathbf p|,
        ~\lambda = \pm\frac32,\pm\frac12,
    \end{gather}
    where the four-spinors are normalized as usual \(u_{\mathbf p\lambda}^\dag u_{\mathbf p\lambda} = 1\).
    The helical states form momentum space monopoles, `hedgehogs' in momentum space,  of charges $\pm 3, \pm1$
    (for information on momentum space topology see \cite{Volovik:2011kg,Volovik2003}). 
    The Berry connections and curvatures are defined as 
    $ \mathcal{A}^{(\lambda)}(\mathbf p)= -iu^{\dagger}_{\lambda\mathbf p}\nabla_{p}u_{\lambda\mathbf p}$ 
    and $\mathbf \Omega^{(\lambda)}(\mathbf p)=\mathbf  \nabla_{p} \times \mathcal A^{(\lambda)}$,
    The curvatures are regular everywhere except the RSW node.
    Monopole charge for a mode \(\lambda\) is equal to the Berry curvature flux through a surface that embraces the zero-energy point.
    Chern number of the node is the sum of monopole charges over the positive-energy modes
\be
    &&
    \mathbf\Omega^{(\lambda)} = \frac{\lambda}{p^3}\mathbf p,\label{Chern}
    \quad N^{(\lambda)}=\int{\frac{d^{3}p}{2\pi} {\nabla_{\mathbf p}\cdot\mathbf\Omega^{(\lambda)}}}=2\lambda,
    \quad\mathcal C_\frac32 = N^{(\frac32)} +N^{(\frac12)} = 4.
\ee
    The Berry curvature is related to the momentum space topological invariants.
    Chern number can be expressed via quasiparticle propagator
\be
    \mathcal C \equiv \mathcal N_3 = \frac{1}{24\pi^2}\tr\int GdG^{-1}\wedge GdG^{-1}\wedge GdG^{-1},\quad G^{-1}=i\omega- H,
\ee
    where the integration is over a 3d-surface (in 4d Euclidean space \((\omega,\mathbf p)\)) that embraces the propagator poles.

    According to the Nielsen-Ninomiya theorem \cite{Nielsen:1983rb}, the Weyl nodes emerge in pairs of opposite chiralities
    (RSW node of the opposite chirality would have low-energy Hamiltonian \eqref{eq1} with the opposite overall sign and \(\mathcal C_{-\frac32}=-4\)).
    Band structure of a real crystal may be quite complicated and highly asymmetric,
    e.g. in a `left-handed' PdGa chiral crystal, an RSW node \(s=\frac32\) with \(\mathcal C_\frac32=4\), which is a fourfold level crossing, 
    is opposed with two coinciding \(s=1\) nodes with \(\mathcal C_{-1}=-2\) each (the sign in front of spin representation corresponds to the overall sign of the low-energy Hamiltonian \eqref{eq1}), which is a sixfold level crossing \cite{Schroter:2020}.
    The generalization of the Nielsen-Ninomiya theorem is the condition on sum of Chern numbers of all nodes of a crystal
    \be
    &&0=\sum_n\mathcal C_n,
    \ee
    where \(n\) labels the nodes.

    The following consideration is for an isolated node.
    Of course, the total vector current gets compensated \(\mathbf j\sim\sum_n\mathcal C_n=0\),  while the axial current gets doubled \(\mathbf j^5\sim\sum_n|\mathcal C_n|\).

    In the rest of this section we provide a quasiclassical and an exact calculation of the CSE current.
    As an exact solution, we introduce a rather general proof of CSE to be a topologically protected effect.
    The quasiclassics is illustrative because it involves a simple yet quite unusual Hamiltonian system.
    The quasiclassical treatment helps us to prove the final result, and is also potentially useful, 
    since various perturbations may be analyzed straightforwardly by solving the kinetic equation with a non-trivial collision integral.

\subsection{Quasiclassical treatment}

    The quasiclassics is known to yield exact results for oscillator-like problems, one of which is the free motion in uniform magnetic field.
    Formally, the method is justified if many Landau levels are occupied, 
    which is the case at the limit of weak magnetic field as compared to the chemical potential, \(\mu^2\gg B\).

    In this section we follow \cite{eqBerry, Son, CSE2}.
    We write the kinetic equation for the quasiparticles using the quasiclassical equations of motion.
    Integration of the kinetic equation over the momentum space yields the continuity equation.
    Then we extract the expression for the current density from the continuity equation.

    The quasiclassical equations of motion in the present case differ from the standard  equations of motion in a uniform magnetic field by the term with the Berry curvature (for a concise derivation, see Appendix \ref{App1})
\be
    &&\dot{\mathbf r}^{(\lambda)}=\frac{\partial E_{\mathbf p\lambda}}{\partial \mathbf{p}}+\dot{\mathbf p}^{(\lambda)} \times \mathbf{\Omega}^{(\lambda)}\label{eqBc}\\
    &&\dot{\mathbf p}^{(\lambda)}=\dot{\mathbf r}^{(\lambda)} \times \mathbf{B} \label{eqBc2}.
\ee

    The kinetic equation  for the quasiparticles reads 
\be
&&\frac{\partial n_{\mathbf{p}\lambda}}{\partial t }+\frac{\partial n_{\mathbf{p}\lambda}}{\partial \mathbf{r}}\mathbf{\dot{r}}^{(\lambda)}+\frac{\partial n_{\mathbf{p}\lambda}}{\partial \mathbf{p}}\mathbf{\dot{p}}^{(\lambda)}=\sum_{\lambda'}I^{(\lambda,\lambda')}_{col} \label{kin}
\ee
    In this, the simplest, model, the fermions of different (quasi)helicities \(\lambda\) do not mix, 
    \(I^{(\lambda,\lambda')}=I^{(\lambda)}\delta_{\lambda\lambda'}\), 
    so the kinetic equation decouples.

    With the equations of motion, we write the kinetic equation to the first order in \(B\) 
	\be
    &\frac{\partial n_{\mathbf{p}\lambda}}{\partial t}+(1+\mathbf{B\Omega}^{(\lambda)})^{-1}
    &\left(\left(\frac{\partial E_{\mathbf{p}\lambda}}{\partial \mathbf{p}} \times \mathbf{B}\right)\frac{\partial n_{\mathbf{p}\lambda}}{\partial \mathbf{p}}
    +\right.\nonumber\\
    &&\left.+\left(\frac{\partial E_{\mathbf{p}\lambda}}{\partial \mathbf{p}}+\left(\frac{\partial E_{\mathbf{p}\lambda}}{\partial \mathbf{p}} \cdot\mathbf \Omega^{(\lambda)}\right)\mathbf{B}\right)\frac{\partial n_{\mathbf{p}\lambda}}{\partial \mathbf{r}}\right)=I^{(\lambda)}_{col} \label{kin1}
	\ee

    The integration of the last equation over the momentum space of the system yields the continuity equation.
    However, the system \eqref{eqBc}, \eqref{eqBc2} does not preserve the standard phase space flow.
    The proper phase space that restores the Liouville's theorem is \cite{Berryphase}
\be
&&d^{3}pd^{3}x \to (1+\mathbf{\Omega^{(\lambda)} B})d^{3}pd^{3}x.
\ee

    Finally, after an integration of \eqref{kin1} over \(\frac{{d^{3}p}}{(2\pi)^{3}}(1+\mathbf{\Omega^{(\lambda)} B})\), we obtain the continuity equation
\be
&&\frac{\partial \rho^{(\lambda)}}{\partial t }+ \mathbf{\nabla}\mathbf{j}^{(\lambda)}=\int{\frac{{d^{3}p}}{(2\pi)^{3}}(1+\mathbf{\Omega^{(\lambda)} B}) I^{(\lambda)}_{col}}\label{eqkinet},\quad 
\ee
    where the desired expression for the current density reads
\be
&& \mathbf{j}^{(\lambda)}=\int\frac{d^{3}p}{(2\pi)^{3}}\left(\frac{\partial E_{\mathbf{p}\lambda}}{\partial \mathbf{p}}+\left(\mathbf{\Omega}^{(\lambda)}\cdot \frac{\partial E_{\mathbf{p}\lambda}}{\partial \mathbf{p}}\right)\mathbf{B}\right)n_{\mathbf{p}\lambda} \label{current}
\ee
    The first term vanishes due to the isotropic dispersion law. 
With the curvature \eqref{Chern} and the Fermi-Dirac distribution
\[n_{\mathbf p\lambda}=\left(\exp(\beta(E_{\mathbf p\lambda}-\mu))+1\right)^{-1},
\] 
    we obtain
\be
&&\mathbf{j}^{(\lambda)}=\frac{N^{(\lambda)}}{4\pi^{2}}\mu\mathbf{B}.
\ee
The total current is the sum over \(\lambda\)
\be
&&\mathbf{j}=\frac{N^{(\frac12)}+N^{(\frac32)}}{4\pi^{2}}\mu\mathbf{B} = \frac{\mu\mathbf B}{\pi^2}.
\ee

Note that the reasoning of this subsection is valid for any \(s\), and the sum over monopole charges in momentum space in the last equation adds up into the corresponding Chern number \(\mathcal C_s\).
Thus, we write CSE current for a node that is described with a low-energy effective theory of form \eqref{eq1} with arbitrary (half-integer) \(s\)
\be
&&\mathbf{j_s}= \frac{\mathcal C_s}{4\pi^2}~\mu\mathbf B.\label{QuasiChernSE}
\ee

\subsection{Exact treatment}
\label{CSEexact}

A problem of free motion in uniform magnetic field is exactly solvable following Landau \cite{Land}.
With the spectra obtained, any observable may be computed.
But the CSE conductivity happens to be a special quantity that is dictated by the topological properties of the spectra, and is robust under a wide class of the theory deformations.
In this subsection we show that the conductivity is expressed via the spectral asymmetry, which is defined by number and type of gapless modes, of the theory.

In the standard gauge $A_{x}=By$ (mind that the magnetic field is directed against the axis, \(B_z=-B\)), we express the Hamiltonian via the creation and annihilation operators
\be
&&a=\frac{\hat{p}_{x}+By-i\hat{p}_{y}}{\sqrt{2B}},
\quad a^\dag=\frac{\hat{p}_{x}+By+i\hat{p}_{y}}{\sqrt{2B}},
\quad S_\pm = S_x \pm iS_y ,\\
&&H=v_{f}\sqrt{\frac{B}{2}} (S_{+}a+S_{-}a^\dag) + v_{f}p_{z}S_{z} = H_\bot + p S_z, \quad p = v_fp_z\label{eqland}. 
\ee

Though the exact spectra is cumbersome, few spectral properties of the Hamiltonian allow to calculate the Chiral Separation current \(j_z\).
We propose the ansatz for the spectral problem
\be
&&H\ket{\psi}=E(p)\ket{\psi}, \quad \braket{\psi_\al(p)|\psi_\beta(p')}=2\pi\delta(p-p')\delta_{\al\beta},\\
&&\ket{\psi} = \ket{p}(c_1\ket{n-3}, c_2\ket{n-2}, c_3\ket{n-1}, c_4\ket{n})^T,\quad C(p)^\dag C(p)=1,
\ee
where \(C(p)=(c_1,c_2,c_3,c_4)^T\) are not constant, and \(\ket{n}\) are the standard oscillator eigenstates (\(a\ket{n}=\sqrt{n}\ket{n-1},~a\ket{0}=0,~\braket{n|m}=\delta_{nm},~\ket{n<0}=0\)).

For a given mode \((E(p), C(p))\) 
\be
&&HC=EC, \quad C^\dag H=EC^\dag,\quad \frac{d}{dp}H_\bot=0,\\
E &=& C^\dag H_\bot C + p~C^\dag S_zC,\\
1 &=& \frac{dp}{dE}~C^\dag S_zC + \frac{dC^\dag}{dE}HC + C^\dag H\frac{dC}{dE} = \\
  &=& \frac{dp}{dE}~C^\dag S_zC + E\frac{d(C^\dag C)}{dE} = \\
  &=& \frac{dp}{dE}~C^\dag S_zC.\label{SpectProp1}
\ee

Another neat property of the spectra for an odd half-integer \(s\) is the asymptotic behavior of the energy bands.
The nonzero matrix elements of \(H_\bot\) for a given band are of order \(\sim n\sqrt{B}\),
while the longitudinal component is unbounded 
\be
&&0 = \det(H_\bot + S_zp_z - E(p_z)) = p_z^4~\det\left(\frac{1}{p_z}H_\bot + S_z - \frac{E(p_z)}{p_z}\right),
\ee
which means (note that it is not the case for an integer \(s\) since \(\hat S_z\ket{s=0,1,...; ~s_z=0}=0\))
\be
&&|E(p_z)|\sim|p_z|\text{ at }~p_z\to\infty \label{SpectProp2}.
\ee
Such energy bands \(E_\al(p_z)\) (\(\al\) labels the modes) essentially fall in three classes based on the type of mapping \(E(p_z): \mathbb R\to\mathbb R\) as follows
\begin{enumerate}
    \item \(E(-\infty)E(+\infty)>0\) --- the band crosses the zero energy level an even number of times (including zero times). We assign to the mode a value \(\Theta_\al=0\);
    \item \(E(-\infty)=-\infty\) and \(E(+\infty)=+\infty\) --- the band crosses zero an odd number of times, and the mapping preserves the orientation, so \(\Theta_\al=1\);
    \item \(E(-\infty)=+\infty\) and \(E(+\infty)=-\infty\) --- the band crosses zero an odd number of times, but the orientation is flipped, so \(\Theta_\al=-1\).
\end{enumerate}

The quantity of interest is straightforwardly expressed via the spectra and the Fermi-Dirac distribution, which is applicable for positive and negative energy solutions
\be
j^{z}&=&\int \frac{B}{2\pi}\frac{dp_z}{2\pi}\sum_{\al}~n_f(E_\al)~v_{f}\braket{\psi_\al|S_{z}|\psi_\al},\label{j5def}\\
n_f(E)&=&\frac{\text{sign}(E)}{\exp(\beta(E-\mu)\text{sign}(E))+1},\quad\beta=T^{-1}.
\ee
Since the transverse motion of a free particle in magnetic field is quantized \cite{Land}, 
the phase space associated with the transverse motion is provided by the factor \(\frac{B}{2\pi}\), where \(2\pi\) is the magnetic flux quantum for the unit charge.

With some rearrangements of \eqref{j5def} we make a use of the first spectral property \eqref{SpectProp1}
\be
j^z&=& \frac{B}{4\pi^2}\int_{-\infty}^{+\infty}dp \sum_{\al}~n_f(E_\al(p))~C_\al^\dag S_{z}C_\al\\
&=& \frac{B}{4\pi^2}\sum_{\al}\int_{E_\al(-\infty)}^{E_\al(+\infty)} dE_\al~n_f(E_\al)~\frac{dp}{dE_\al} ~C_\al^\dag S_{z}C_\al\\
&=& \frac{B}{4\pi^2}\sum_{\al}\int_{E_\al(-\infty)}^{E_\al(+\infty)} dE~n_f(E).
\ee
The problem with the integral is if the integration over the band energy values spans from \(-\infty\) to \(+\infty\),
which is dictated by the second spectral property \eqref{SpectProp2}. 

Convergence of the integral over \(E_\al\) specifies, what is `infinite energy' in context of the second spectral property: 
\(|E_\al|\) must at least exceed \(|\mu|+T\).
E.g.~for PdGa chiral crystals \cite{Schroter:2020}, the depth of the Dirac sea (near both nodes) is about \(0.5\) eV, which constrains the theory for \(\mu<0\) (room temperature is about \(0.3\) eV).

After some rearrangements of the last integral, and with a remarkable property of the standard Fermi-Dirac distribution \(n^{-1}(E)=1+\exp(\beta E)\)
\be
&&\int_0^{+\infty} dE{(n(E-\mu)-n(E+\mu))}=\mu \label{distdiff},
\ee
we obtain
\be j^z &=& \frac{\mu B}{4\pi^2}\sum_{\al}~\Theta_\al.\label{topCSE} \ee

The fact that CSE conductivity is expressed via spectral asymmetry \(\sum_\al\Theta_\al\) exhibits topological nature of the effect.
The result \eqref{topCSE} holds for a `deformed' theory that possess spectral properties \eqref{SpectProp1} and \eqref{SpectProp2}.
Even more, the first spectral property is fair for a half-integer \(s\), 
while the second spectral property is fair for an odd half-integer \(s\), but can be generalized to integer \(s\) 
(\(\Theta\)'s in that case take values \(0,\pm\frac12,\pm1\); 
~\(\pm\frac12\) is for bands that interpolate between 0 and \(\infty\) --- two such bands of the same sign are, in a sense, one band that crosses zero once at infinity; 
another peculiarity is that, for instance, the \(s=1\) spectra contains a countable set of zero modes with \(\Theta_{n0}=0\) that cross zero twice --- at \(p_z=0\) and at infinity). 

To obtain the final answer, we just count the spectral asymmetry \(\sum_\al\Theta_\al\) --- sum up the values \(\Theta_\al\) for the modes, which cross the zero energy level odd number of times.
A band that crosses zero odd number of times does it at least once.
Let us list the gapless modes at the zero energy points and its \(\Theta\)'s
\be
\ket{\psi_0} &=& \ket{p_z=0}(0,0,0,\ket{0})^T,~\Theta_0=-1,\\
\ket{\psi_{1\pm}} &=& \ket{p_z=\pm 2\sqrt{2B}}(0,0,\pm\frac{\sqrt3}{2}\ket{0},\frac12\ket{1})^T,~\Theta_{1\pm}=-1,\\
\ket{\psi_2} &=& \ket{p_z=0}(0,\sqrt{\frac35}\ket{0},0,\sqrt{\frac25}\ket{2})^T,~\Theta_2=-1.
\ee
The first one, \(\ket{\psi_0}\), is a simple mode with linear dispersion \(E=-\frac{3}{2} v_{f}p_{z}\).
It is not obvious if \(\ket{\psi_{1\pm}}\) are the zeroes of two distinct modes, or two zeroes of one mode.
The dispersion relations for these two modes follow from a quadratic equation, so we checked with brute force that these are distinct modes.
An exact computation of the dispersion relation for the last one, \(\ket{\psi_2}\), would be quite involved. 
Fortunately, we don't need it. 
Recalling the magnetic field direction (\(B_z=-B\)), we finally obtain
\be
j^z &=& \frac{\mu B}{4\pi^2}\sum_{\al=0,1+,1-,2}(-1) = -\frac{\mu B}{\pi^2},\quad
\mathbf j = \frac{\mu \mathbf B}{\pi^2}.
\ee

We noted that the spectral asymmetry of the Hamiltonian in magnetic field is equal to the node Chern number 
(the minus sign is due to the magnetic field direction \(B_z=-B\))
\be 
\sum_{\al}-\Theta_\al=\mathcal C_s,\label{LandauIndexEq} 
\quad \mathcal C_{\pm\frac12}=\pm1,~\mathcal C_{\pm1}=\pm2,~\mathcal C_{\pm\frac32}=\pm4,~ ...
\ee 
(the sign in front of the spin representation \(s\) corresponds to the overall sign of the low-energy effective Hamiltonian of form \eqref{eq1})
which, apparently, is a manifestation of an index theorem.
The relation was recently proved in \cite{LandauIndex}, index theorems for Dirac operators in open spaces were studied in context of axial anomaly, 
and spectral asymmetry of Dirac operators --- in context of fractional fermion numbers.

Meanwhile, we have already proved the relation \eqref{LandauIndexEq} implicitly.
Firstly, the spectral asymmetry \(\sum_\al\Theta_\al\) is the same for any \(B>0\).
Secondly, the quasiclassical result of the previous section \eqref{QuasiChernSE} is accurate for small enough \(\mu\) and \(B\), \(\mu^2\gg{B}\).
Thirdly, scope of application of the exact result \eqref{topCSE} covers the scope of quasiclassics.
Hereby, cross-linking of the exact and quasiclassical results yields the desired relation \eqref{LandauIndexEq}.

Thus, we find the CSE conductivity for nodes that have low-energy effective theory of form \eqref{eq1} for any \(s=\frac12,1,\frac32,...\) 
(mind that \(s\) is quasispin --- regardless its value, all the quasiparticles obey Fermi statistics)
\be
\sigma_s = \frac{\mu}{4\pi^2}\mathcal C_s \label{ChernSE},
\ee
where \(\mathcal C_s\) is Chern number of the node. 
We explicitly checked the result for \(s=\frac12,~1,~\frac32\).

For a crystal with a set of Fermi points, we consider the whole Brillouin zone and find that the total CSE current reads
\be
\mathbf j_5 = \mathbf B \frac{\mu}{4\pi^2}\sum_n|\mathcal C_n|.
\ee
The absolute values in the formula are also justified with symmetry arguments.
\(\mathbf j_5\) and \(\mathbf B\) are P-odd, which requires P-even CSE conductivity, while \(\mathcal C\)'s itself are P-odd.

\end{section}

\begin{section}{Discussions}
    \label{SectConc}

We studied CSE in systems with various momentum space topological properties: RSA model, which is a relativistic theory of combined spin 3/2 and 1/2 massless fields, and models of topological semimetals with multi-degenerate Fermi-points.
CSE in chiral systems similar to Weyl fermions is defined by their Chern numbers, which are easily read of helicities of the plane-wave solutions.
The simple rule applied to a complex system, like RSA model, helps to understand the structure of CSE (and, apparently, of the chiral anomaly), which is quite unclear from the perturbative calculation.

For RSA model, as for other examples, the coefficients in CSE and in the chiral anomaly coincide.
We obtained the static CSE conductivity using the Kubo formula, which involves the one-loop polarization operator.
A relation between the triangle diagrams and the polarization operator was suggested in \cite{Adler1,Adler2,Adler3}.
With that relation, equivalence of the coefficients might be proved.
However, the proof would be quite cumbersome.
In our perturbative calculations, we used the strong coupling limit \(m\to\infty\) and circumvented the problem of ghost degrees of freedom,
which made the result look gauge- and parameter-dependent.
A way to comprehensively study the anomaly and CSE in RSA model is to calculate the anomaly by means of dispersion relations \cite{VDZK,DZ1971,DZK,CutRul1,CutRul2}.
Such calculation would also demonstrate that the effects emerge at IR as well as at UV.
We failed to prove in a general case, or find in the literature, the expression for the topological invariant \(\mathcal N_3\) for a chiral system via helicities of the (gapless) modes.
The expression would imply scale-independence of CSE and the anomaly, and validity of our result for any \(m>0\).

Starting with a calculation of CSE for a model of a four-fold band crossing in RSW semimetals, we studied the effect for a class of models of multi-fold band crossings in topological semimetals.
The effect is topologically protected --- it is robust to deformations that preserve the Chern numbers of the Fermi points, e.g.~the effect holds for anisotropic crystals.
However, the result has certain limitations,
e.g.~the theory may fail for negative chemical potential due to ``shallowness of the Dirac sea'' near a Fermi point in a real semimetal.
Unclear, if the effect survives a vacuum rearrangement, e.g.~transition to superconductive state.
Besides, CSE lacks a sound  experimental manifestation so far; may be the effect contributes to magnetic properties of topological semimetals, or yields unusual properties of a semimetal-metal contact.

An index theorem that we accidentally discovered is formulated for open space operators.
All other open space index theorems that we could find in the literature are reduced to the Atiyah–Singer index theorem by some compactification, none of which seems applicable in the present case.
A view of the problem from mathematical standpoint would be illuminating.

\end{section}

\begin{section}{Acknowledgements}
    The authors thank M.A.Zubkov  for useful discussions and strong support at every stage of the work. 
    The work was supported by Russian Science Foundation Grant No. 16-12-10059.
\end{section}

\begin{appendices} \begin{section}{}\label{App1} In this section, we briefly derive semiclassical equations of motion in a crystal following \cite{eqBerry}

An electron wave packet is defined as
\be
    &&\ket{\Psi}=\int d^{3}q~ a(\mathbf{q},t)\ket{\psi(\mathbf{x}_{c},t)},
    \quad\ket{\psi_{\mathbf{q}}(\mathbf{x}_{c},t)}=\exp(iq\hat{x})\ket{u({\mathbf{x}_{c}},\mathbf{q},t)}\\
&&\int{d^{3}q}|a(\mathbf{q},t)|^{2}=1,\quad
\braket{\psi_{\mathbf{q}}|\psi_\mathbf{q^{'}}}=\delta(\mathbf{q}-\mathbf{q^{'}}), \\
&&\mathbf{q}_{c}=\int{d^{3}q~\mathbf{q}|a(\mathbf{q},t)|^{2},\quad
    \mathbf{x}_{c}=\braket{\Psi|\hat{x}|\Psi}},\quad
    a(\mathbf{q},t)=|a(\mathbf{q},t)|\exp(-i\gamma(\mathbf{q},t)).
\ee
    The amplitude change is to be negligible at interatomic distances for validity of the quasiclassical description.

The Lagrangian for an electron wave packet is defined as
\be
&&L=\braket{\Psi|i\frac{d}{d t}-\hat{H}|\Psi},
\ee
which yields
\be
&&\braket{\Psi|i\frac{d}{dt}\Psi}=\frac{d\gamma}{dt}-\dot{\mathbf{q}}_{c}\cdot \mathbf{x}_{c}+\dot{\mathbf{q}}_{c}\cdot\braket{u|i\frac{\partial u}{\partial\mathbf{q}_{c}}}+\dot{\mathbf{x}}_{c}\cdot\braket{u|i\frac{\partial u}{\partial\mathbf{x}_{c}}}+\braket{u|i\frac{\partial u}{\partial t}}\\
&&\frac{d\gamma}{dt}=\frac{\partial \gamma}{\partial t}+\mathbf{\dot{q}}_{c}\cdot\frac{\partial\gamma}{\partial t}\\
&&E(\mathbf{x}_{c},\mathbf{q}_{c},t)=\braket{\Psi|\hat{H}|\Psi}
\ee 
    The magnetic field \(\mathbf{B}=\nabla\times\mathbf{A}\) is introduced with the covariant derivative $\mathbf{q} \to \mathbf{q}+\mathbf{A}(\mathbf{x}_{c})$.
    So we  obtain  the Lagrangian
\be
&&L=-E+\frac{d(\gamma-\mathbf{x}_{c}\cdot\mathbf{\mathbf{\dot{q}_{c}})}}{dt}+\dot{\mathbf{q}}_{c}\cdot\braket{u|i\frac{\partial u}{\partial\mathbf{q}_{c}}}+\dot{\mathbf{x}}_{c}\cdot\braket{u|i\frac{\partial u}{\partial\mathbf{x}_{c}}}+\braket{u|i\frac{\partial u}{\partial t}}+\mathbf{q}_{c}\cdot \mathbf{ \dot{x}}_{c}.
\ee
    The corresponding equations of motion are \eqref{eqBc} and \eqref{eqBc2}.
\end{section}

\end{appendices}

\end{document}